\RequirePackage{fix-cm}
\documentclass[twocolumn]{svjour3}          
\smartqed  
\usepackage{graphicx}
\newcommand{\styofa}{Sr$_2$(Mg$_y$Ti$_{1-y}$)O$_3$FeAs}
\newcommand{\svofa}{Sr$_2$VO$_3$FeAs}
\newcommand{\stofa}{Sr$_2$TiO$_3$FeAs}
\newcommand{\stmofa}{Sr$_2$Ti$_{\frac{1}{2}}$Mg$_{\frac{1}{2}}$O$_3$FeAs}
 \journalname{J Supercond Nov Magn}
\begin{document}

\title{Effects of metallic spacer in layered superconducting \styofa
}


\author{Kwan-Woo Lee         
}


\institute{K.-W. Lee \at
 Department of Display and Semiconductor Physics,
 Korea University, Sejong 339-700, Korea\\
 Department of Applied Physics, Graduate School,
 Korea University, Sejong 339-700, Korea\\
              \email{mckwan@korea.ac.kr}           
}

\date{Received: date / Accepted: date}

\maketitle

\begin{abstract}
The highly two-dimensional superconducting system \styofa, recently synthesized 
in the range of $0.2 \le y \le 0.5$, shows an Mg concentration-dependent $T_c$.
Reducing the Mg concentration from $y$=0.5 leads to a sudden increase in 
$T_c$, with a maximum $T_c \approx$40 K at $y$=0.2.
Using first principles calculations, the unsynthesized stoichiometric $y$=0
and the substoichiometric $y$=0.5 compounds have been investigated.
For the 50\% Mg-doped phase ($y$=0.5), Sr$_2$(Mg$_y$Ti$_{1-y}$)O$_3$ layers 
are completely insulating spacers between FeAs layers,
leading to the fermiology such as that found for other Fe pnictides.
At $y$=0, representing a phase with metallic Sr$_2$TiO$_3$ layers,
the $\Gamma$-centered Fe-derived Fermi surfaces (FSs) considerably shrink 
or disappear.
Instead, three $\Gamma$-centered Ti FSs appear, and in particular
two of them have similar size, like in MgB$_2$.
Interestingly, FSs have very low Fermi velocity in large fractions:
the lowest being 0.6$\times$10$^6$ cm/s.
Furthermore, our fixed spin moment calculations suggest the possibility of 
magnetic ordering, with magnetic Ti and nearly nonmagnetic Fe ions.
These results indicate a crucial role of Sr$_2$(Mg$_y$Ti$_{1-y}$)O$_3$ layers 
in this superconductivity.
\end{abstract}

\section{Introduction}
\label{intro}
Since Hosono and coworkers discovered superconductivity in hole-doped LaFeAsO
at $T_c$=26 K,\cite{hosono} Fe-pnictides have been the cause of much excitement.\cite{ishida}
The mechanism of the superconductivity has yet to be uncovered,
but many have discussed the nesting of two Fermi surfaces at the $\Gamma$ 
and $M$ points.\cite{mazin_prl08,zhiping,mazin_nat09,bang09}
However, some experimental and theoretical investigations 
of a few superconducting Fe-pnictides
have shown the absence of a nested fermiology, though these compounds are expected 
to share a common mechanism of the superconductivity. \cite{LP_svofa,arita,bori,hding}

In the pnictide family, the recently synthesized 21322 systems 
Sr$_2$${\cal M}$O$_3$Fe${Pn}$ (${\cal M}$=transition metal, and ${Pn}$=P or As),
with a thick perovskite-like spacer between Fe$Pn$ layers,
have stimulated a great enthusiasm due to its possible bimetallic character 
as well as its high two-dimensionality.\cite{ogino,zhu,sato,ogino1,ogino2,ogino3}
Sato {\it et al.} synthesized a member of the 21311 system \styofa~ 
in the range of $0.2 \le y \le 0.5$.\cite{sato} With 50\% Mg doping of the Ti sites ({\it i.e.}, 
$y$=0.5), the superconductivity appears at $T_c$= 10 K, 
but with a small superconducting volume fraction, 
which thus implies that there is no intrinsic superconductivity.
On reducing the Mg concentration, $T_c$ increases sharply to $\sim$ 34 K
at $y=0.45$, and reaches a maximum $\sim$ 40 K at $y$=0.2,
slightly higher than that in a stoichiometric \svofa.
Throughout all of the doping range considered, 
no significant changes in the lattice constants were observed.
Although synthesis of a sample below $y$=0.2 has not been
successful yet, the Ti rich phase may be expected to have a higher $T_c$.
This implies that the Ti ions play a crucial role in the superconductivity
of this system.

From the viewpoint of formal charge, the $y$=0.5 phase ({\it i.e.}, 50\% Mg doping) 
consists of tetravalent Ti ions of $d^0$, 
leading to a wholly insulating block between FeAs layers.
Replacing a fraction of the sites containing Mg ions with Ti ions, 
Ti$^{4+}$ ion becomes metallic $d^{1-2y}$ Ti$^{(3+2y)+}$ on the average.
Thus, this system is a good example for investigating the effects of metallic and
insulating spacers on the superconductivity.
The effects of a metallic spacer have been under debate for \svofa~ 
having V ($d^2$) layers.\cite{LP_svofa,mazin_prb10}

In this paper, we investigate the $y$=0.5 and 0 phases, which represent either
pure insulating or metallic spacers, respectively.
So far, no evidence of magnetic ordering has been reported in this compound. 
We therefore will focus on the electronic structures and Fermi surfaces of 
the nonmagnetic (NM) state, 
though possible magnetic tendencies will also be discussed briefly.

\begin{figure}[tbp]
{\resizebox{5.5cm}{10cm}{\includegraphics{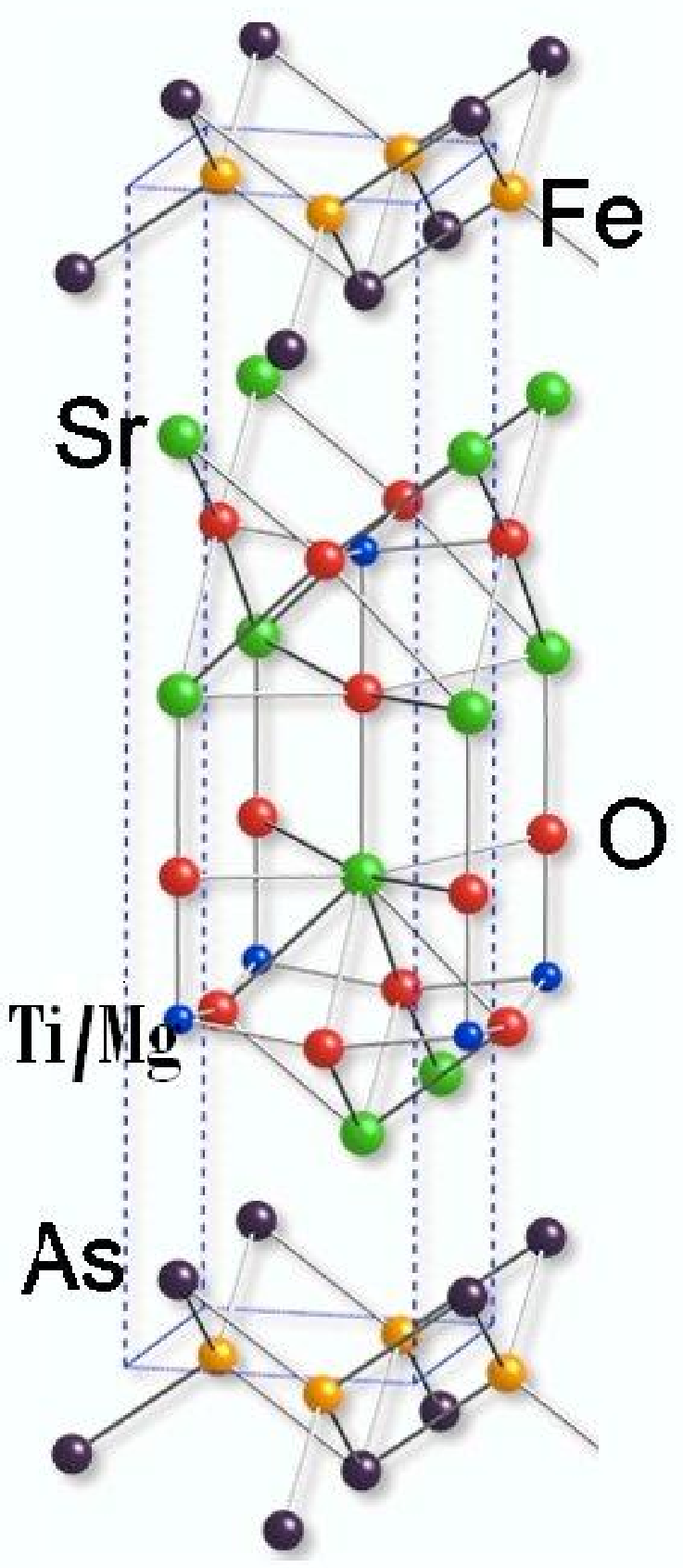}}}
\caption{Crystal structure of \styofa, in which FeAs layers are separated by two
Sr$_2$$\cal{M}$O$_3$ blocks. The Fe-As bond length is 2.32 \AA.
}
\label{str}
\end{figure}

\section{Structure and Calculation Method}
In our calculations, based on the tetragonal unit cell with $P4/nmm$ space group,
the experimentally measured lattice constants
$a$=3.935 \AA~ and $c$=15.952 \AA~ were used,
resulting in a slightly larger volume by 2\% than in \svofa.\cite{sato}
In the cell, Fe atoms lie at $2a$ sites ($\frac{1}{4}$,$\frac{3}{4}$,0),
Ti/Mg, Sr, As, and O atoms lie at $2c$ sites ($\frac{1}{4}$,$\frac{1}{4}$,$z$),
and another O atoms sit at the $4f$ sites ($\frac{1}{4}$,$\frac{3}{4}$,$z$).
Using the local density approximation (LDA),
the internal parameters were optimized to
0.3072 for Ti, 0.8210 and 0.5898 for two Sr sites, 0.0772 for As,
0.4270 for O at $2c$ sites, and 0.2939 for O at $4f$ sites.
These values lead to the As-Fe-As bond angle of $\alpha$=116$^\circ$,
which is measured in one side of the Fe layer, 
corresponding to $T_c \approx$25 in the literature.\cite{clee08}
This significant difference from $T_c \approx$40 K 
suggests a crucial role of the Sr$_2$$\cal{M}$O$_3$ layer.
The Mg-rich phase \stmofa~ was understood with a $2\times1$ supercell
using our optimized internal parameters. 
In these optimization processes, the residual Hellman-Feynman forces
on the atoms were lower than 1 meV/\AA.

In all calculations, the local spin density approximation (LSDA) implemented
in the accurate all-electron full-potential
local orbital code, FPLO, was used.\cite{fplo1}
A regular mesh containing 196 $k$ points in the irreducible
wedge was used to sample the Brillouin zone.
The magnetic tendencies in the $y$=0 phase were studied
using the fixed spin moment (FSM) method,\cite{fsm} with a much denser
mesh of up to 726 irreducible $k$ points.

\section{Results}

\begin{figure}[tbp]
\vskip 8mm
{\resizebox{8cm}{6cm}{\includegraphics{Fig2a.eps}}}
\vskip 12mm
{\resizebox{8cm}{6cm}{\includegraphics{Fig2b.eps}}}
\caption{Top: Band structure, with
the fatband of Ti $t_{2g}$ manifold, of nonmagnetic
2(\stmofa)~ in the $2\times1$ supercell.
The Ti $d_{zx}$ band lying around 1.8 eV is nearly dispersionless
and separated from the Ti $d_{yz}$ band.
Bottom: Enlarged fatband representation of Fe $d$ bands
near the Fermi energy $E_F$, which is set to zero. 
The character of Fe $d_{xy}$ band appears in the range of 
--2.2 to -1.1 eV and 0.6 to 1.6 eV (not shown here).
In the fatband representation, the size of the symbols is proportional to
the fractional character of each orbital.
}
\label{band21}
\end{figure}

\begin{figure}[tbp]
\vskip 8mm
{\resizebox{8cm}{6cm}{\includegraphics{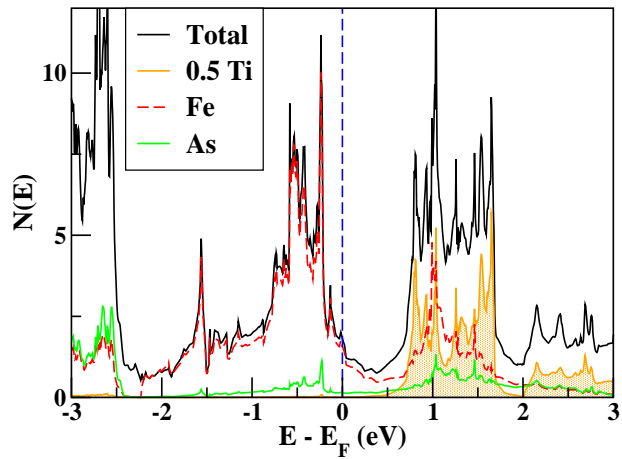}}}
\caption{Total and atom-projected densities of states (DOSs)
on nonmagnetic \stmofa~ in the $2\times1$ supercell.
DOS at $E_F$ $N(E_F)$ is 1.82 states per eV per formula unit, 30\% smaller
than the value in LaFeAsO.\cite{singh_prl08}
}
\label{dos21}
\end{figure}

\subsection{50\% Mg doped phase: \stmofa}
First, we will address the electronic structure of the $y$=0.5 phase
having the insulating spacer. 
Figure \ref{band21} displays the enlarged band structure with the fatband of
the Ti $t_{2g}$ manifold (top panel) and the fatband of the Fe $d$ manifold (bottom panel).
Note that the $X$ and $M$ points in the unit cell are folded into the
$\Gamma$ and $X$ points in the $2\times 1$ supercell, respectively.
The corresponding DOS near the Fermi energy $E_F$ is given in Fig. \ref{dos21}.
As expected, in the $y$=0.5 phase, the $3d$ orbitals of the Ti ions 
are completely unoccupied. The Ti $t_{2g}$ manifold expands from 0.5 eV to 2 eV.
In this phase, the details near $E_F$ are very similar to those observed for 
other superconducting Fe-pnictides.\cite{singh_prl08,LP_fese,lee_nafeas,mazin_prb08}
At the $\Gamma$ point, three hole pockets derived from the Fe $d_{x^2-y^2}$, $d_{yz}$, 
and $d_{zx}$ bands appear, while two-fold $M$-centered electron pockets with 
the character of these three bands also appear.
Considering that there is no intrinsic superconductivity at this phase and 
the similar fermiology with other superconducting Fe pnictides, 
the nesting effects may not be a necessary ingredient for Fe pnictides 
to be superconductors.

\begin{figure}[tbp]
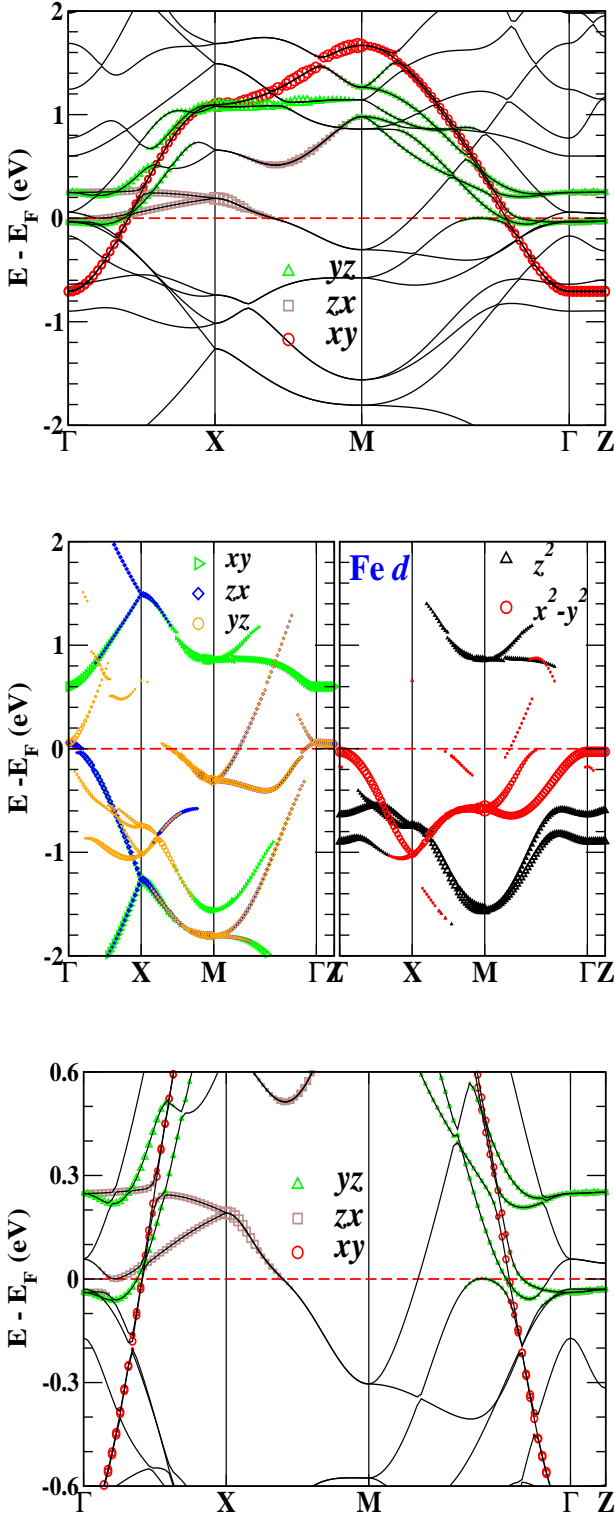

\vskip 8mm
{\resizebox{8cm}{6cm}{\includegraphics{Fig4a.eps}}}
\vskip 10mm
{\resizebox{8cm}{6cm}{\includegraphics{Fig4b.eps}}}
\vskip 10mm
{\resizebox{8cm}{6cm}{\includegraphics{Fig4c.eps}}}
\caption{Top: Band structure,
overlapped with the fatband of Ti $t_{2g}$ manifold, on nonmagnetic \stofa.
Middle: Fatband representation of Fe $d$ bands.
Bottom: Enlarged band structure near $E_F$ with the fatband of Ti $t_{2g}$ manifold,
which shows interesting features (see text).
}
\label{band}
\end{figure}

\begin{figure}[tbp]
\vskip 8mm
{\resizebox{8cm}{6cm}{\includegraphics{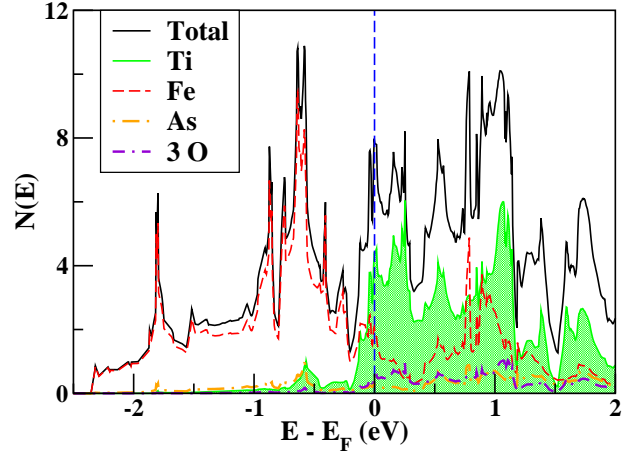}}}
\caption{Total and atom-projected densities
of states (DOSs), near $E_F$, on nonmagnetic \stofa.
$E_F$, denoted by the vertical dashed line, lies on
a sharp peak.
DOS $N(E_F)$ at $E_F$  is 8.0 states/eV per formula unit:
55\% Ti, 25\% Fe, and 5\% for each O.
}
\label{dos}
\end{figure}

\subsection{Electronic structure of nonmagnetic $y$=0 phase}
Now, we will focus on the $y$=0 phase, which contains only 
the metallic $d^1$ Ti$^{3+}$ ions.
Thus, this represents the phase of the range of $0 \le y <0.5$, 
consisting of metallic spacers.
In this subsection, we will address the electronic structure of the NM state.
The enlarged band structure near $E_F$  and the fatband of the Ti $t_{2g}$
manifold are displayed in the top panel of Fig. \ref{band}.
The incomplete TiO$_5$As octahedron results in a breaking of the symmetry 
of the $t_{2g}$ manifold.
The partially filled Ti $d_{xy}$ band spreads over the range of
--0.7 to 1.6 eV (here, $E_F$ is set to zero),
which is a width that is 15\% larger than that of the V $d_{xy}$ band in \svofa.\cite{LP_svofa}
This band can be described by a single-band tight-binding
model with nearest neighbor hopping of $t$=0.28 eV and next neighbor hopping of
$t'$=0.08 eV, {\it i.e.}, values that are about 15\% larger than in \svofa.\cite{LP_svofa}
The other bands of Ti are mostly unoccupied, but
there is some $E_F$ crossing in the $d_{yz}$ and $d_{zx}$ bands
near the $\Gamma$ point, as shown in the bottom panel of Fig. \ref{band}.
Although the hybridization of FeAs layers and the intervening Ti layers
is negligible in almost the whole regime due to high two-dimensionality,
small mixing with Fe $d_{zx}$ and $d_{yz}$ leads to this $E_F$ crossing (see below).

Compared with other superconducting Fe-pnictides, \cite{singh_prl08,LP_fese,lee_nafeas,mazin_prb08}
the bands having Fe character, as displayed in the middle panel of
Fig. \ref{band}, show both similarities and remarkable distinctions.
Similar to other superconducting Fe pnictides,
the $d_{yz}$ and $d_{zx}$ bands lead to electron pockets at the $M$ point. 
However, instead of three Fe-derived hole pockets at the $\Gamma$ point, 
only one Fe $d_{zx}$-derived hole pocket appears, but this is considerably shrunk 
in this system compared to other Fe-pnictides (see below).
This may imply that the role of Fe for the superconductivity is substantially 
reduced in this phase.

Most of the Fe bands are separated from the Ti-derived bands,
reflecting strong two-dimensionality.
Exceptions occur around the $M$-point and near the $\Gamma$ point, near $E_F$.
Near the $\Gamma$ point, the Fe and Ti $d_{zx}$ bands are hybridized 
with each other, leading to a 0.1 eV gap at $E_F$.
On the other hand, around the $M$ point, the Ti $d_{zx}$ are more strongly
hybridized with the Fe $d_{zx}$ and $d_{yz}$, with the gap of 0.6 eV 
that occurs above $E_F$, resulting in surviving $M$-centered electron pockets.
This difference to \svofa~ results from the different $d$-orbital filling
between V and Ti ions in these systems: namely $d^2$ for V$^{3+}$ and $d^1$ for
Ti$^{3+}$.

As shown in the bottom panel of Fig. \ref{band}, noticeable features appear
close to $E_F$.
At the $\Gamma$ point, two-fold bands at --40 meV and 58 meV, and
a single band at --30 meV appear.
One of the two-fold valence bands is dispersionless near the $\Gamma$
point along the $\Gamma$--$X$ line, leading to a van Hove singularity
at -40 meV.
On the other hand, one of the doublet conduction bands
almost touches with $E_F$ along the $\Gamma$--$X$ line, 
leading to a van Hove singularity almost exactly at $E_F$.
Additionally, one band lying on $E_F$ appears along the $M$--$\Gamma$ line
(actually a few meV above $E_F$).

These features are reflected in the total and atom-projected densities of states
(emphasized near $E_F$), given in Fig. \ref{dos}.
A valley appears at -20 meV between the two van Hove singularities 
at -40 meV and $E_F$, indicating sensitivity to hole (or electron) doping.
This may imply that the stoichiometric sample is near to the optimal doping.
$N(E_F)$ is twice as large than for most other Fe pnictides  
due to the greater contribution of Ti, 
but is only two thirds of that found in \svofa.\cite{LP_svofa}
Remarkably, except for the Ti contribution, 
the magnitude of $N(E_F)$ is similar with that of LaFeAsO.
It is interesting that the $T_c$ of hole-doped LaFeAsO is comparable with the $y=0.5$
system described herein, in which Ti ions are insulating.
This fact suggests that metallic Ti ions can play an important role 
in the superconductivity of this system.

\subsection{Fermi surface of nonmagnetic $y$=0 phase}
\begin{figure}[tbp]
\resizebox{4.5cm}{4cm}{\includegraphics{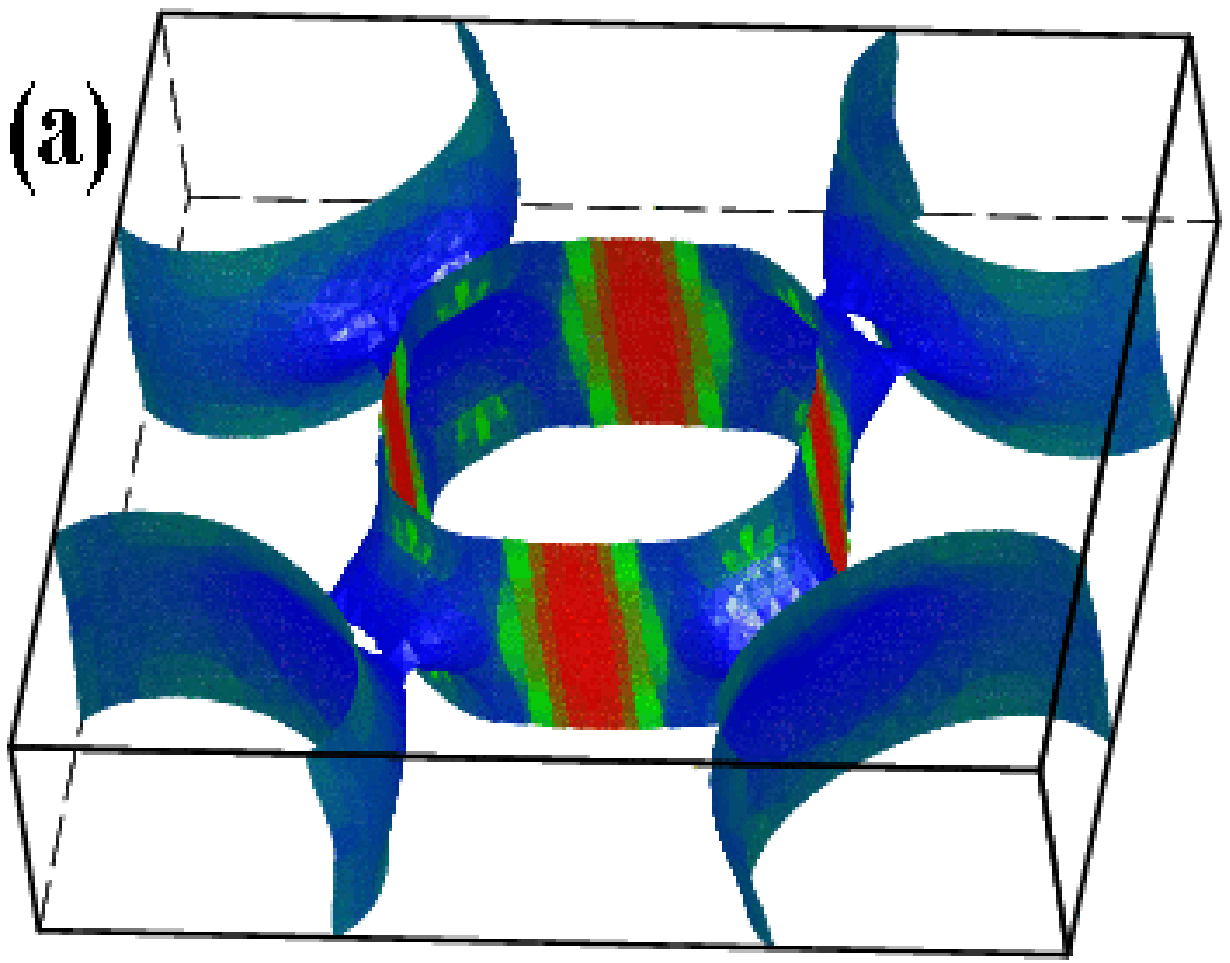}}
\resizebox{4.5cm}{4cm}{\includegraphics{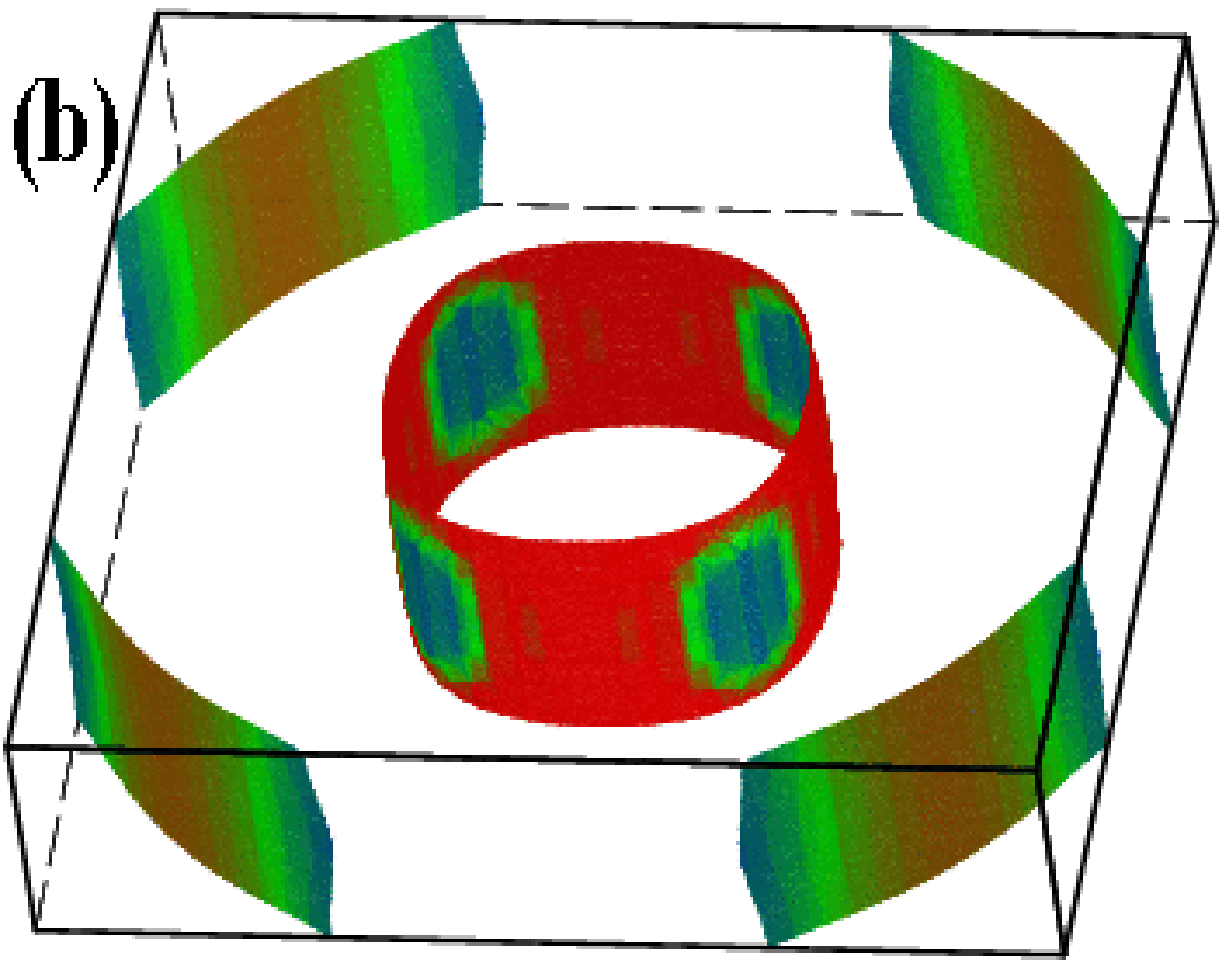}}
\resizebox{6.2cm}{5cm}{\includegraphics{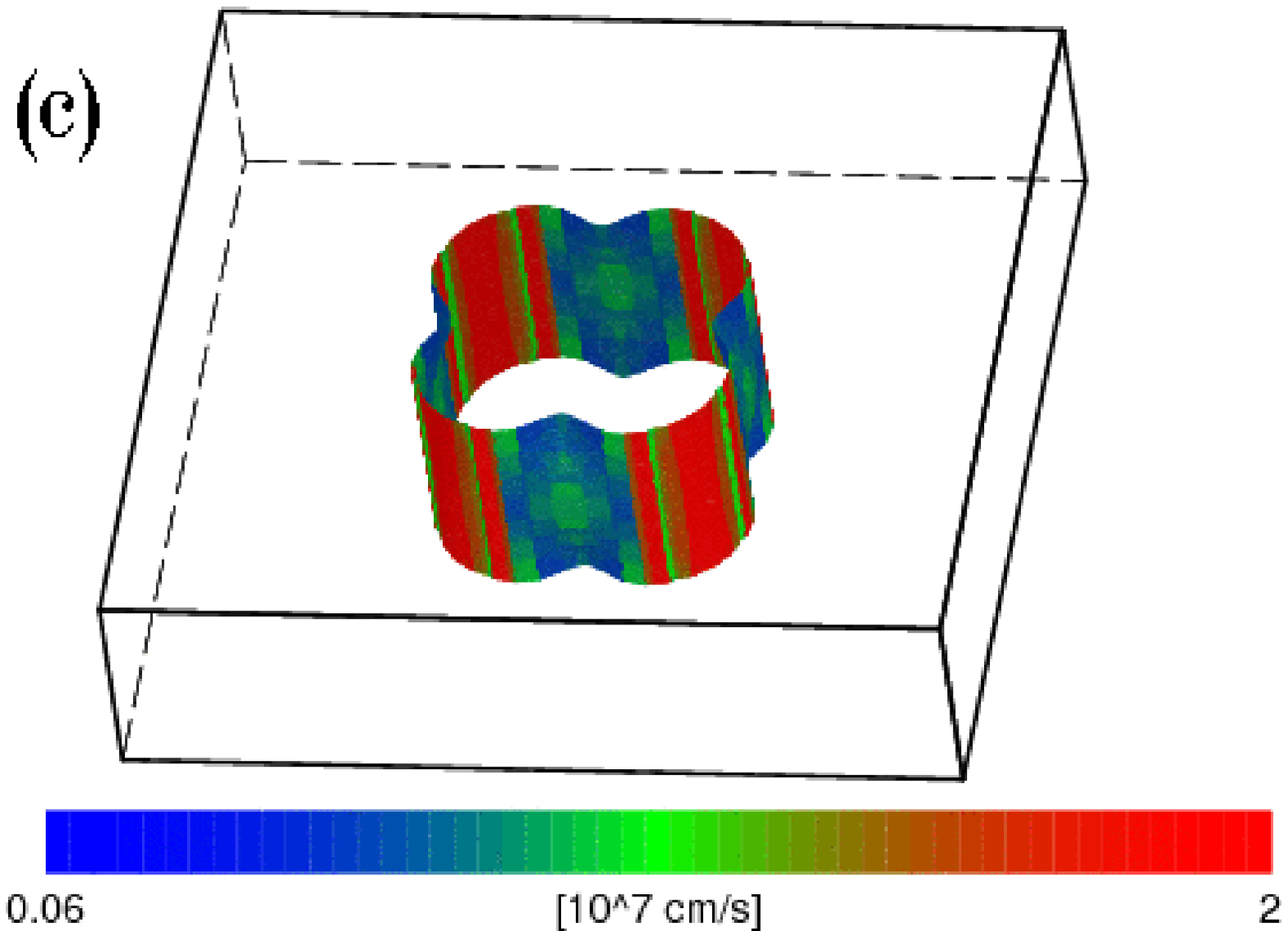}}
\resizebox{4.5cm}{4cm}{\includegraphics{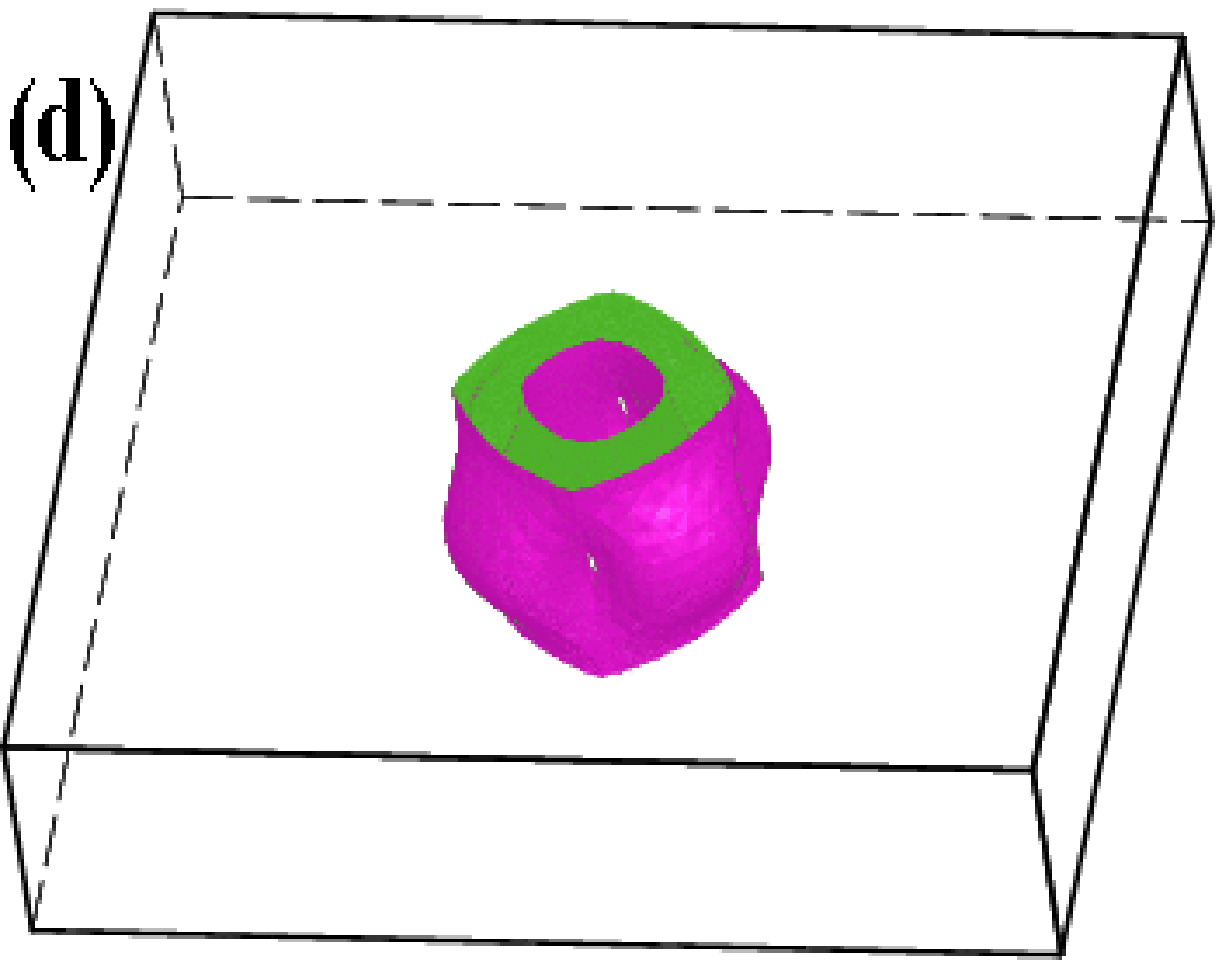}}
\caption{Fermi surfaces (FSs) of nonmagnetic \stofa.
(a)--(c) Fermi velocities $v_F$ colored dark (blue) for the lowest
$0.06\times10^7$ and lighter (red) for the highest $2\times10^7$,
in units of cm/s.
Throughout FS of (d), $v_F$ is remarkably low,
0.6 -- 7$\times10^6$ cm/s.
Only (d) and the $M$-centered FSs have Fe character.
}
\label{fs}
\end{figure}

The Fermi surfaces (FSs) consist of five $\Gamma$-
and two $M$-centered pockets, as illustrated in Fig. \ref{fs}.
Three large $\Gamma$-centered electron pockets have Ti character 
and the others have Fe character, derived from the $d_{zx}$ band.
The character of the $M$-centered FSs is mostly derived from the
$d_{zx}$ and $d_{yz}$ bands and partially from the $d_{x^2-y^2}$ bands.
The $M$-centered electron pockets possess the shape of 
an elliptical cylinder (in Fig. \ref{fs}(a)) and a rhombus (in Fig. \ref{fs}(b)).
In the regime between the two Fe-derived $\Gamma$-centered FSs, 
electrons reside in a coaxial distorted rectangular box-like shape 
(see Fig. \ref{fs}(d)).
A Fe-derived $\Gamma$-centered hole pocket is considered to be a crucial 
ingredient for the formation of pairings in Fe pnictides,
but in this system the pocket is shrunk considerably, 
implying the opposing view that the metallic Ti ions have a more important role.
The three $\Gamma$-centered electron pockets are derived from
Ti $d_{xy}$, $d_{yz}$, and $d_{zx}$, see Figs. \ref{fs}(a) to (c).
The first two FSs, which are of similar size, are much like MgB$_2$.

A large portion of the FSs has remarkably low Fermi velocity $v_F$,
with the lowest being $0.6\times10^6$ cm/s. In particular, 
throughout the $M$-centered elliptical cylinder and the $\Gamma$-centered 
concentric Fe-derived FSs, $v_F$ is close to its lowest value.
For the Ti-derived $\Gamma$-centered FSs, 
$v_F$ is quite low along the $\Gamma - M$ line for the two similar sized 
FSs and along the $\Gamma - X$ line in the remaining FS given in Fig. \ref{fs}(c).

\subsection{Magnetic tendencies of the $y$=0 phase}
In this subsection, we will address briefly the possible magnetic tendencies
in the Mg-free phase.
The competition between superconductivity and magnetic ordering
widely observed in Fe pnictides is unclear for this compound,
since no magnetic ordering has been observed.
Allowing for a ferromagnetic (FM) state, FM has an energy that is a little lower
by 5 meV/Fe than for NM. The total magnetic moment is 0.46 $\mu_B$/Fe,
mostly comprised of contributions from the Ti ions (0.37 $\mu_B$).
The Fe ions are nearly nonmagnetic, less than 0.1 $\mu_B$,\cite{unrelaxed}
suggesting that the magnetic fluctuation due to Ti ions is of
more importance than that due to the Fe ions in this phase.

The fixed spin moment (FSM) method was employed to investigate
this magnetic behavior, using a cell that allows only FM and NM states.
As obtained from LSDA calculations, FSM shows the minimum energy
at a total fixed moment of $M$=0.45 $\mu_B$/Fe,
with an energy 6 meV/Fe lower than in the NM state.
From $M$=0 to around the minimum energy state, the Fe moment is
quite small and most of the contribution to $M$ results from the Ti moment.
After the minimum energy is reached, the moment of Fe increases monotonically,
coinciding with a sharp increment in energy with a slope of $\sim$ 220 meV/$M$
($M$: total moment per formula unit), while that of Ti stays at
around 0.6($\pm$0.05) $\mu_B$.
The Stoner $I$ is 0.47 eV, so $IN(E_F)$=2.2 with $N(E_F)$=4.6 states per eV
per formula unit obtained from the FM calculations, indicating strong
magnetic instability.

\section{Discussion and Summary}
For the $y$=0 phase, the on-site Coulomb repulsion $U$ to Ti ions
was applied using both popular double-counting schemes 
in the LDA+U approach,\cite{ldau1,ldau2} 
to investigate whether or not a Mott transition in the $d^1$ Ti$^{3+}$ ions occurs.
Up to $U$=7 eV $\sim$ $3W$, which is a much larger value than 
what has been used in perovskite ${\cal R}$TiO$_3$ (${\cal R}$=rare earth elements),\cite{eva04} 
the Sr$_2$TiO$_3$ layers are still metallic.
This failure of LDA+U has previously been observed in some $4d$ or $5d$ systems, 
in which no integer occupation number can occur due to strong $p$-$d$ hybridization,
since an integer occupation is required to lead to a Mott transition by the LDA+U 
method.\cite{LP_BNOO} 
In this system, the calculated occupation number is significantly larger than 
the formal number $d^1$, resulting from the absence of rigorous definition
of oxidation state in solids, as recently discussed.\cite{sit,jiang}
This may prevent the LDA+U approach from driving a Mott transition.\cite{pickett}

In summary, we addressed the electronic structure of \styofa~ 
for the $y$=0.5 and $y$=0 phases, which represent insulating and metallic Ti layers, 
respectively. At $y$=0.5, which shows no intrinsic superconductivity, 
the fermiology is just like that for other Fe-pnictides. 
Introducing metallic Ti ions, the $\Gamma$-centered
Fe-derived FSs are reduced considerably or disappeared.
Instead, three $\Gamma$-centered Ti FSs appear, and two of them have similar
sizes like in MgB$_2$. 
Our FSM calculations further suggest possible magnetic ordering in the Ti ions. 
Although no magnetic ordering has been observed in this compound, 
the high temperature resistivity measurement shows a kink
for samples below $y$=0.5 at $T \sim 200$ K,\cite{sato} 
thus implying magnetic transition.
Consequently, a detailed magnetic measurement is required for this compound.
Our results indicate that the metallic Ti ions play a crucial role 
in the superconductivity, suggesting that the mechanism of superconductivity
in Fe-pnictides should be reconsidered.


\begin{acknowledgements}
We acknowledge W. E. Pickett for fruitful communications and H. Ogino 
for clarifying his experimental results on superconductivity of the $y$=0.5
phase.
This research was supported by NRF of Korea under Grant No. 2012-0002245.
\end{acknowledgements}


\begin{thebibliography}{10}
\bibitem{unrelaxed} Using the same internal parameters
as in SVOFA\cite{LP_svofa} and the experimental lattice parameters
in this system\cite{sato},
FM has the large Fe moment of 2.16 $\mu_B$, resulting in total moment
2.53 $\mu_B$, but much higher energy by $\sim$ 315 meV/f.u. than in the state
with nearly nonmagnetic Fe ions which is obtained with our relaxed
internal parameters. This fact implies strong magneto-phonon coupling
on Fe ions in this system as observed in LaFeAsO.\cite{zhiping}
\end{thebibliography}


\end{document}